# Genetic Algorithm: Reviews, Implementations, and Applications

Tanweer Alam(✉)
Faculty of Computer and Information Systems, Islamic University of Madinah, Saudi Arabia
`tanweer03@iu.edu.sa`

Shamimul Qamar
Computer Engineering Department, King Khalid University, Abha, Saudi Arabia
Amit Dixit
Department of ECE, Quantum School of Technology, Roorkee, India
Mohamed Benaida
Faculty of Computer and Information Systems, Islamic University of Madinah, Saudi Arabia

**Abstract**—Nowadays genetic algorithm (GA) is greatly used in engineering pedagogy as an adaptive technique to learn and solve complex problems and issues. It is a meta-heuristic approach that is used to solve hybrid computation challenges. GA utilizes selection, crossover, and mutation operators to effectively manage the searching system strategy. This algorithm is derived from natural selection and genetics concepts. GA is an intelligent use of random search supported with historical data to contribute the search in an area of the improved outcome within a coverage framework. Such algorithms are widely used for maintaining high-quality reactions to optimize issues and problems investigation. These techniques are recognized to be somewhat of a statistical investigation process to search for a suitable solution or prevent an accurate strategy for challenges in optimization or searches. These techniques have been produced from natural selection or genetics principles. For random testing, historical information is provided with intelligent enslavement to continue moving the search out from the area of improved features for processing of the outcomes. It is a category of heuristics of evolutionary history using behavioral science-influenced methods like an annuity, gene, preference, or combination (sometimes refers to as hybridization). This method seemed to be a valuable tool to find solutions for problems optimization. In this paper, the author has explored the GAs, its role in engineering pedagogies, and the emerging areas where it is using, and its implementation.

# 1 Introduction

The GA is a meta-heuristic motivated by the evolution process and belongs to the large class of evolutionary algorithms in informatics and computational mathematics. These algorithms are frequently used to create high-quality solutions to optimize and search concerns by focusing on bio-inspired operators such as selection, convergence, or mutations [1]. The author John Holland developed GAs based on Darwin's evolutionary theory in 1988 [2]. Subsequently, in 1992, he expanded the GA [3]. This algorithm falls under the heading of evolutionary algorithms. The evolutionary algorithms are used to solve problems that do not already have a well-defined efficient solution. This approach is used to solve optimization problems (scheduling, shortest path, etc.), and in modeling and simulation where randomness function is used [4]. GA is a solution to the population of the candidate (known as people, animals, or genotypes) to the problem of optimizing that is developed towards better options [5]. Every candidate's solution has a set of characteristics (the genes or phenotype) that can be evolved and changed; typically, solutions are depicted in the binary digits as strings of 0s and 1s, although another codec is also allowed. Evolution generally starts of a community of randomized individuals and is an iterative process with the population being viewed as a method of generation for each reproduction. For every generations, the fitness of everyone in the population is measured. However, the fitness is usually the value of the objective feature being solved [6]. When sufficiently fit individuals are probabilistically chosen from the existing population, and the gene is modified to create a new generation cycle for all (recombined and potentially mutated at random) [7]. A newer generation of candidate strategies would be utilized over the next generation of the process. The algorithm usually ends when either a maximum number of generations or satisfaction has been generated. Therefore, every successive generation is more suitable for the environments of the population [8]. Within the search's approaches, the populations is maintained. Every person representing the solutions to a provided issue in the computational complexity. Everyone in the population is numbered as a finite length vectors of components that they have [9]. The component is like Genes and many genes generate a chromosome. The fitness score is represented to everyone that has the ability of an individual to fulfill. An optimal fitness score could be found for the individual [10]. GA can maintain the population of n persons along with the fitness scores that they have. Everyone is having good fitness score that are given more chance to reproduce. Figure 1 shows Gene, Chromosome, and population. Any individual that has good fitness score is selected whose mating and generate good offspring by grouping chromosomes of their generation. When a new baby born, the room will be created since the population size is static. Thus, several persons expire and to get replaced with new arrivals that ultimately create a new generation if all the older human population breeding potentials are low. Whether the less suitable expire, there is a possibility that alternative solutions would be sought across succeeding generations. Over avg, these newer generations provide more "good genes" than that of the older generation's person. Therefore, every new generation has good solution than the old generation. When the offspring generated with no significant differences than offspring generated by the old population, the populations are converging. This algorithm known as converted to group of solution for the problem individually. Following are the strengths of GAs.
1. The GA is robust and strong.
2. It provides an optimistic solution over large populations.

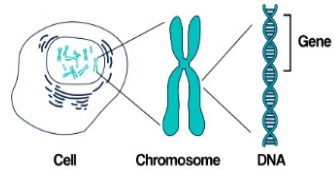

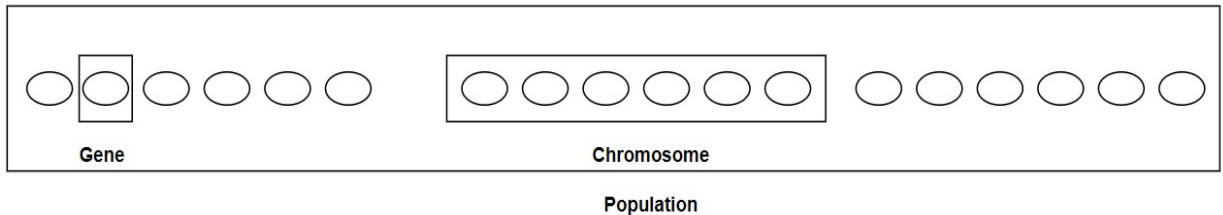

**Fig. 1.** Gene, Chromosome, and population

3. GAs do not break on slight changes in input or noise present during the process.

The research questions are: "How is the GA useful in engineering pedagogy?", "How can we implement GA in emerging areas?".

The rest of the paper is organized as below. Section II represents the GA operators, Section III represents the GA, Section IV shows the methodologies of this algorithm, Section V represents the role of GA in different areas and section VI shows the GA applications, section VII represents the conclusion.

## 2 Operators used in GA

When the initial generation is generated, the GA evolves the generation using the operators, discussed as follows.

### 2.1 Operator for Selection

In this kind of operator, the individuals have a preference with better fitness score and enable them to pass their gene to the successive generation of the individual.

### 2.2 Operator for Crossover

This reflects a generation of breeding among persons. Randomly choose two individuals using selection operators and crossover operators. After chosen, genes are at the crossover sites exchanged, so building a completely new offspring or individual. Figure 2 shows the crossover operator.

### 2.3 Operator for Mutation

To use this tool, introduce random genes into offspring to retain the genetic heterogeneity to prevent excessive divergences. Figure 3 shows the mutation process.

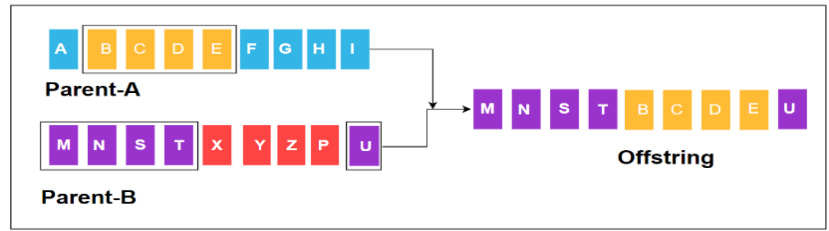

**Fig. 2.** Crossover operator

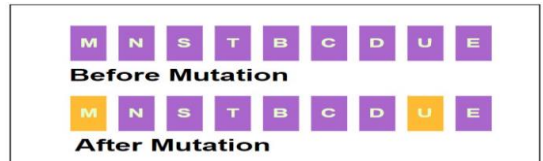

**Fig. 3.** Mutation

## 3 Genetic algorithm

It is a paradigm of machine learning that generates the behavior patterns from the representation of evolution mechanisms. It is achieved through generation on the inside of the machine of the population of entities identified via genetics. People in the population will undergo a phase of mutation. It should be noted that growth will not be an assisted mechanism. Therefore, there is no evidence to support the theory that the purpose of evolution is to generate humans. So, the mechanisms of existence appear to come away to various Persons competing for services in the World. The following steps are used to obtain fitness using GA (Figure 4).
1. Consider populations p randomly.
2. Obtain the fitness of the population.
3. Repeat from Step 4 to 7 until convergence.
4. Choose any parent from the population individually.
5. Generates a new population through the crossover process.
6. Insert random genes in a new population to perform mutation.
7. Obtain fitness for newly generated populations.
The whole process is described as follows.
Suppose there is a target string, its goal is to produce target string starting from a random string with the same length. Its process for implementation is as follows.
1.	The characters A-Z, a-z, 0-9, and other special symbols are considered as genes in GA.
2.	The string generated by these characters is considered a chromosome.

The fitness score is the number of characters that are differed from characters in target string at a specific index of the string. Therefore, individuals having lower fitness value is given more preferences. According to the output of the algorithm, this approach having issues in the optimal solutions so that further improvement is needed to update the fitness score.

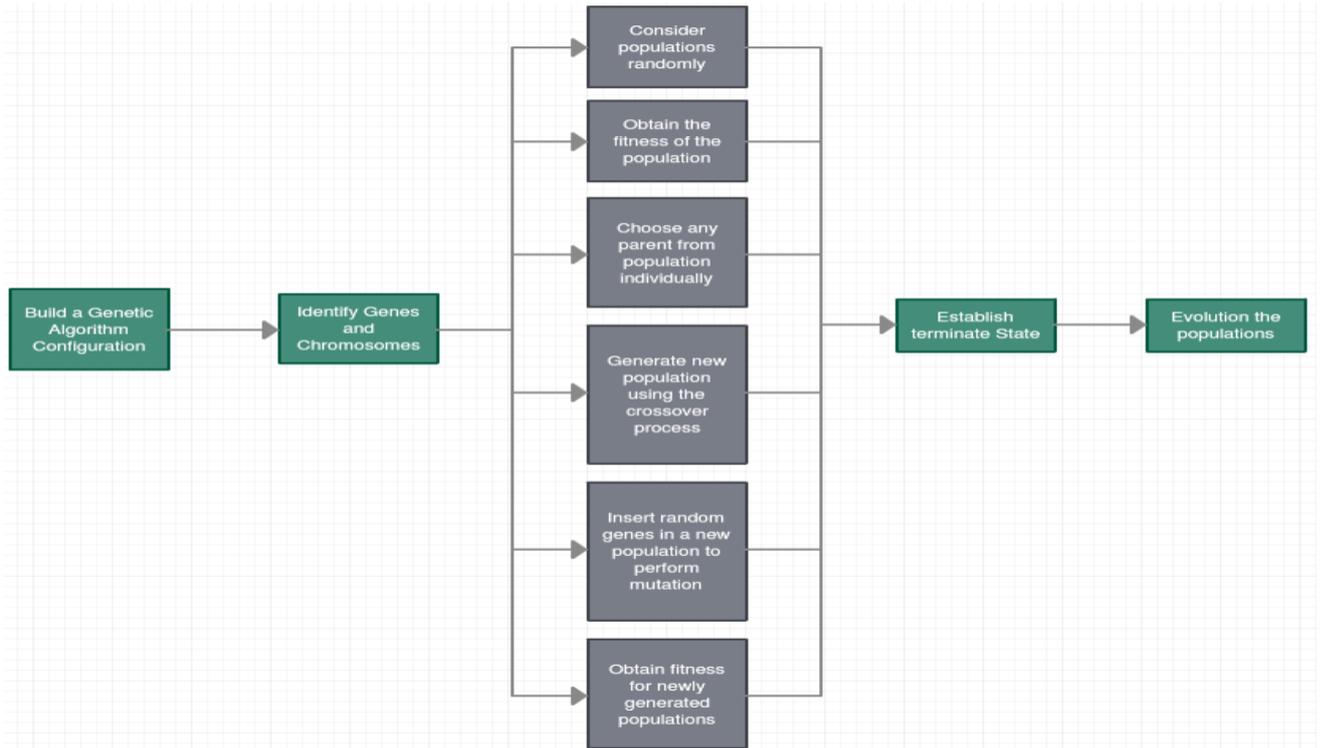

**Fig. 4.** Genetic Algorithm

## 4 Methodologies

### 4.1 Generate New Population

Firstly, throughout order to create new communities, several individual strategies are randomly generated. The population density is calculated as per the scope of the problem. Usually, moreover, it includes many hundreds or thousands of potential solutions for maximizing certain population numbers. The populations are produced typically at random, providing the whole range of potential alternatives. Furthermore, ideas in places where appropriate approaches are expected to be produced that can be "seeded".

### 4.2 Selections

The proportion of the current population is chosen at the period of each subsequent generation to breed a new-generation process. To use the health, feature the individual solution is chosen via a health-based procedure. The basic approaches of evaluation assess the health for each outcome choose better solution. Other approaches also score a random sample of the populations, as this approach can be very time-consuming. All the mechanisms are stochastic and built to choose a limited proportion of approaches that are less fit. This helps to maintain population heterogeneity high and to avoid excessive convergence on a poor response.

### 4.3 Reproductions

It is an approach to regenerate a generation population of solution from selected using genetic operator such as crossovers and mutations operators. Increasing new solutions to be generated is a couple of parental strategies that are chosen from the randomized collection for breeding. The reproduction process generates a child solution using crossover and mutation operators that shares the characteristics of the parent solution. Every new child has a new parent and this process will continue until the new population generated with the expected sizes. The new generation population generated with different chromosomes. Therefore, the average fitness is increased by this procedure for the populations. Figure 5 shows the methodologies used in GAs.

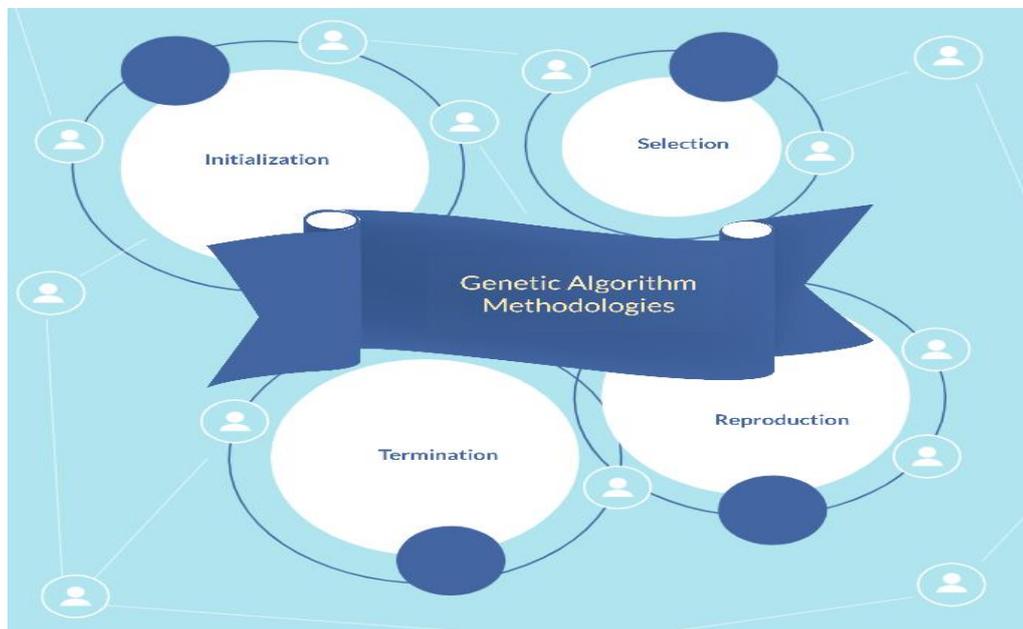

**Fig. 5.** Methodologies

**4.4    Termination**

The production process will continue until a termination condition has occurred. The Genetic experiments are carried out on chosen people. That outcomes of the genetic activities are incorporated into the new population of the peoples. Whether the termination requirement is met throughout this stage, the current population will continue to be replaced by the new population and Stage 2-4 will be continued. Therefore, the best individual results of each population are identified. The following terminating conditions are occurring:

1. Generated solution satisfies the minimum criteria.
2. The expected size of populations generated.
3. The computational requirements fulfilled such as time or money.
4. Best fitness solutions found.
5. Physical evaluation.

**5    Genetic Programming**

The genetic programming is the main application of GA. A result of the GA is an amount, whereas the result of genetic programming is a virtual machine process. Essentially, this would be the starting of the algorithms that run automatically. But many researchers think how genetic programming and GAs are entirely different things according to their features. Furthermore, the genetic programming is fundamentally very distinct from another methodology to artificial intelligence, machine learning, neural networks, evolutionary structures, deep learning, or computational reasoning because of how it is genetically motivated; it performs its quest for a result of poverty within the framework of development. Ultimately Genetic Programming helps machines to find solutions without any need for programmers to decide the technique, i.e. what can be performed. That performs such an objective of intelligent coding by genetically modifying a population of automated systems based on the concepts of Natural selection Evolutionary theory or biologically inspired activities. Primarily, sequentially, the genetic design turns a population of software programs into something like a newer generation of software by adding additives to biologically active genetic activities. The Genetic processes involve fusion, mutation, replication, chromosome replication, or genome removal. The Genetic programming is ideally suited to many kinds of challenges. So, there's no transportation option in an automobile. Many approaches operate efficiently at the cost of money, whereas others operate quickly at a high overall possibility. Operating a vehicle, thus, entails balancing speeds against protection, and several other factors. Throughout this case, genetic programming would offer a solution that tries to adapt and was the most effective approach from a wide number of factors [41], [42]. To start implementing GA, pursue the following steps:
1. Build a GA Configuration.
2. Identify Genes and Chromosomes.
3. Apply Fitness Method.
4. Establish a terminate State.
4. Evolution of the populations.

## 6 Role of GA in Emerging Areas

The population is configured first. Overall population services are assessed or allocated statistical scores for fitness. The preceding action is taken as long as the current population is completely crowded: human beings from the population are chosen using a selection method. GAs are using in many areas nowadays. Here, the author has highlighted some very important emerging areas as follows.

### 6.1 In Engineering Pedagogy

In the engineering pedagogy, GA is very useful. Some of the areas such as timetable generation, a mathematical model of the problem, optimization principles and practices, etc. At the points of rapid increase as indicated by the growing groups of conference, workshop and publication relating to, and the formation of the journals. The GAs had also considered being a successful method with about its wonderful reliability for several implementations for optimizing, designing, controlling, and machine learning processes. A student who is taking the GA courses and applying the different methods of GA. Also, the evaluation experiments with implementations can play a vital role in the teaching of similar courses in computer science [43].

### 6.2 In the Internet of Things

In the Internet of things (IoT) [15], smart device selection is a challenging task[16], [17], [18]. The GA is the best solution for selecting the IoT nodes to exchange information among IoT nodes [19], [20], [21]. Also, the GA can be used for mining the intelligent objects in IoT [22], [40].

### 6.3 Smart Traffic Signal System

The GA can be implemented in the intelligent traffics. Such an intelligent vehicle sensors scheme was presented by combining IoT with existing highway systems to reduce journey times, emissions of contaminants, traffic, and travelling difficulties. A traffic signals have been constructed throughout every city and rural region at the cross-point of the roadway[23]. Almost all motorcycle, vehicle, bus, auto-rickshaw, transportation providers are for travelling. Utilizing smart signals this could lead to heavy transmissions in the public highway [46].

### 6.4 Intelligent Routing in MANET of Smart Devices

The GA can be implemented in the intelligent routing when smart devices want to discover and connect the neighborhood devices then they need to have an intelligent routing [25]. Nowadays, technological advancement supports various wireless network variants and, between each other, MANET is of utmost significance and a significant challenge. Each important component of MANETs is the geometrical transition with moment. A saving of power generation or movement measured with devices [26] tends to occur at all. Also, as result, the routing problem under MANET is also flexible and could be constructed to discover an optimum path as a dynamic optimization issue. These are important in order to identify the achieving the optimal path the solution must have a high level of capability. The whole objective function accomplished optimum solution routing in a very efficient manner under dynamic topology.

### 6.5    The loads balancing and tasks scheduling in Cloud Computing

GA can be implemented in the loads balancing in the cloud computing system. In the next era, Cloud-MANET can play a vital role in the integration of IoT devices [24]. This integrated framework will be heterogeneous with many technologies.

### 6.6    Blockchain Technology

In blockchain technology, the GA can be implemented. Blockchain on the edge of the network is emerging with a new computing approach for releasing computational resources from access point to 5G edge notes [13],[27]. The blockchain-based processing is being proposed that offloads methods. Similarly, blockchain is an efficient strategy for the distributed network, and an edge computing structure based on blockchain is constructed to diminish the possibility of data failure by incorporating it with blockchain[14].

## 7    GA Implementation in traveling salesman problem

The GA is used to solve the Traveling salesman problem that is a well-known combinatorial problem using novel crossover approaches. Traveling salesman problem is a hybrid computing framework under the broader scope of operation equivalent to the standard transport real-world problems [11]. For applying GA in traveling salesman problems, the encoding mechanism generates a sequence which is then considered a chromosome comprised of a set of items. Such pieces are recognized as genes that make up the chromosomes. The chromosomes are made up of the gene. Selection, crossover, and mutation are the steps of the GAs used in the Traveling Salesman Problems. Chromosomes are the sequence of places reached by salespersons. Consider the following example, S = (S1, S2, S3, ..., Sn) implies that salespersons travel from S1 to S2, S2 to S3, S3 to Sn. There are six places that the salesman will be going through. These places are P1, P2, P3, P4, P5, and P6. This trip starts at P1 and finishes at the same place P1 [12]. Figure 6 indicates the distance between places. The salesman travels according to table 1.

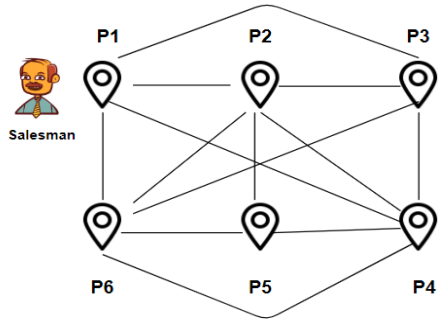

**Fig. 6.** Travelling Salesman Problem

**Table 1.** Travel History

| S. No. | From | To | Distance |
|---|---|---|---|
| 1 | P1 | P2 | 5 |
| 2 | P1 | P3 | 3 |
| 3 | P1 | P4 | 4 |
| 45 | P1 | P5 | 6 |
| 6 | P1 | P6 | 2 |
| 7 | P2 | P3 | 7 |
| 8 | P2 | P4 | 4 |
| 9 | P2 | P5 | 3 |
| 10 | P2 | P6 | 5 |
| 11 | P3 | P4 | 9 |
| 12 | P3 | P5 | 8 |
| 13 | P3 | P6 | 8 |
| 14 | P4 | P5 | 4 |
| 15 | P4 | P6 | 3 |
| 16 | P5 | P6 | 6 |

Apply the GA for the traveling salesman problem scenario.

### 7.1 Obtain the chromosomes.

**Table 2.** Generate the chromosomes

| S. No. | Chromosomes | | | | |
|---|---|---|---|---|---|
| 1 | P2 | P3 | P5 | P4 | P6 |
| 2 | P3 | P2 | P5 | P4 | P6 |
| 3 | P3 | P4 | P6 | P5 | P2 |
| 4 | P4 | P3 | P6 | P5 | P2 |
| 5 | P5 | P2 | P3 | P4 | P6 |
| 6 | P5 | P3 | P6 | P4 | P2 |
| 7 | P6 | P5 | P3 | P4 | P2 |
| 8 | P6 | P4 | P5 | P3 | P2 |

### 7.2 Calculate the initial fitness function.

**Table 3.** Obtain the initial fitness function

| S. No. | Route | Fitness |
|---|---|---|

| 1 | P1P2 | P2P3 | P3P5 | P5P4 | P4P6 | P6P1 |    |
|---|------|------|------|------|------|------|----|
|   | 5    | 7    | 8    | 4    | 3    | 2    | 29 |
| 2 | P1P3 | P3P2 | P2P5 | P5P4 | P4P6 | P6P1 |    |
|   | 3    | 7    | 3    | 4    | 3    | 2    | 22 |
| 3 | P1P3 | P3P4 | P4P6 | P6P5 | P5P2 | P2P1 |    |
|   | 3    | 9    | 3    | 6    | 3    | 5    | 29 |
| 4 | P1P4 | P4P3 | P3P6 | P6P5 | P5P2 | P2P1 |    |
|   | 4    | 9    | 8    | 6    | 3    | 5    | 35 |
| 5 | P1P5 | P5P2 | P2P3 | P3P4 | P4P6 | P6P1 |    |
|   | 6    | 3    | 7    | 9    | 3    | 2    | 30 |
| 6 | P1P5 | P5P3 | P3P6 | P6P4 | P4P2 | P2P1 |    |
|   | 6    | 8    | 8    | 3    | 4    | 5    | 34 |
| 7 | P1P6 | P6P5 | P5P3 | P3P4 | P4P2 | P2P1 |    |
|   | 2    | 6    | 8    | 9    | 4    | 5    | 34 |
| 8 | P1P6 | P6P4 | P4P5 | P5P3 | P3P2 | P2P1 |    |
|   | 2    | 3    | 4    | 8    | 7    | 5    | 29 |

The fitness 29 is reselected that shows the chromosomes with mid value fitness could have a high possibility of being reselected.

### 7.3 Selection

**Table 4.** Obtain value of chromosomes

|   | Chromosome |    | Value    |
|---|------------|----|----------|
| 1 | 1          | 29 | 0.034483 |
| 2 | 1          | 22 | 0.045455 |
| 3 | 1          | 29 | 0.034483 |
| 4 | 1          | 35 | 0.028571 |
| 5 | 1          | 30 | 0.033333 |
| 6 | 1          | 34 | 0.029412 |
| 7 | 1          | 34 | 0.029412 |
| 8 | 1          | 29 | 0.034483 |
| Total |        |    | 0.269631 |

**Table 5.** Calculate Probability

| S. No. | Chromosome value | Total value | Probability  | Cumulative Probability |
|--------|------------------|-------------|--------------|------------------------|
| 1      | 0.034483         | 0.269631    | 0.127889597  | 0.127889597            |
| 2      | 0.045455         | 0.269631    | 0.168582248  | 0.296471845            |
| 3      | 0.034483         | 0.269631    | 0.127889597  | 0.424361442            |
| 4      | 0.028571         | 0.269631    | 0.105963335  | 0.530324777            |
| 5      | 0.033333         | 0.269631    | 0.123624509  | 0.653949286            |
| 6      | 0.029412         | 0.269631    | 0.109082413  | 0.763031699            |

| 7 | 0.029412 | 0.269631 | 0.109082413 | 0.872114112 |
|---|----------|----------|-------------|-------------|
| 8 | 0.034483 | 0.269631 | 0.127889597 | 1.000000000 |

**Table 6.** Generate random number

| S. No. | Random Value |
|--------|--------------|
| 1 | 0.24473 |
| 2 | 0.34523 |
| 3 | 0.65741 |
| 4 | 0.11766 |
| 5 | 0.23123 |
| 6 | 0.54621 |
| 7 | 0.56312 |
| 8 | 0.44344 |

### 7.4 Crossover

The crossover is performed to generate the child from parents who are married. Their resultant chromosomes have been supposed to boost the fitness level. An amount of chromosome that crossover encounter is estimated by the crossover possibility.

**Table 7.** Generate random number after crossover

| S. No. | Random Value |
|--------|--------------|
| 1 | 0.76588 |
| 2 | 0.37643 |
| 3 | 0.98345 |
| 4 | 0.65876 |
| 5 | 0.21543 |
| 6 | 0.23765 |
| 7 | 0.18745 |
| 8 | 0.64398 |

**Table 8.** Generate new chromosomes

| Old | New | Chromosomes | | | | |
|-----|-----|----|----|----|----|----|
| 1 | 4 | P4 | P3 | P6 | P5 | P2 |
| 2 | 1 | P2 | P3 | P5 | P4 | P6 |
| 3 | 7 | P6 | P5 | P3 | P4 | P2 |
| 4 | 8 | P6 | P4 | P5 | P3 | P2 |
| 5 | 5 | P5 | P2 | P3 | P4 | P6 |
| 6 | 6 | P5 | P3 | P6 | P4 | P2 |
| 7 | 2 | P3 | P2 | P5 | P4 | P6 |
| 8 | 3 | P3 | P4 | P6 | P5 | P2 |

### 7.5 Mutation

The mutations mechanism would be brought out during this part. Each mutation acts to swap genes with the other chromosome. The predicted outcomes would increase the importance of fitness to be obtained. When the gene is transferred at the ends of the gene, the mutation will be substituted for the first chromosome. This is a criterion to assess so many chromosomes need to be a mutation.

**Table 9.** New chromosomes after mutation

| S. No. | Chromosomes | | | | |
|---|---|---|---|---|---|
| 1 | P4 | P3 | P6 | P5 | P2 |
| 2 | P2 | P3 | P5 | P4 | P6 |
| 3 | P6 | P5 | P3 | P4 | P2 |
| 4 | P6 | P4 | P5 | P3 | P2 |
| 5 | P5 | P2 | P3 | P4 | P6 |
| 6 | P5 | P3 | P6 | P4 | P2 |
| 7 | P3 | P2 | P5 | P4 | P6 |
| 8 | P3 | P4 | P6 | P5 | P2 |

**Table 10.** Calculate Fitness value after mutation

| S. No. | Route | | | | | | Fitness |
|---|---|---|---|---|---|---|---|
| 1 | P1P4 | P4P2 | P2P6 | P6P5 | P5P3 | P3P1 | |
|  | 4 | 4 | 5 | 6 | 8 | 3 | 30 |
| 2 | P1P2 | P2P5 | P5P3 | P3P6 | P6P4 | P4P1 | |
|  | 5 | 3 | 8 | 8 | 3 | 4 | 31 |
| 3 | P1P6 | P6P4 | P4P5 | P5P3 | P3P2 | P2P1 | |
|  | 2 | 3 | 4 | 8 | 7 | 5 | 29 |
| 4 | P1P6 | P6P4 | P4P2 | P2P3 | P3P5 | P5P1 | |
|  | 2 | 3 | 4 | 7 | 8 | 6 | 30 |
| 5 | P1P5 | P5P3 | P3P2 | P2P6 | P6P4 | P4P1 | |
|  | 6 | 8 | 7 | 5 | 3 | 4 | 33 |
| 6 | P1P5 | P5P6 | P6P3 | P3P2 | P2P4 | P4P1 | |
|  | 6 | 6 | 8 | 7 | 4 | 4 | 35 |
| 7 | P1P3 | P3P5 | P5P2 | P2P4 | P4P6 | P6P1 | |
|  | 3 | 8 | 3 | 4 | 3 | 2 | 23 |
| 8 | P1P3 | P3P6 | P6P4 | P4P2 | P2P5 | P5P1 | |
|  | 3 | 8 | 3 | 4 | 3 | 6 | 27 |

### 7.6 Discussion

It is shown in the first generation that there is the lowest fitness valuation which does not alter. Unless the estimation continues up to the Nth, the smallest fitness level is considered to remain in place. While the measurement is significantly elaborate up to the first generation, a close-optimal approach has been identified from the GA method mentioned, an outcome of the route with the smallest optimum route is P1, P2, P3, P4, P5, P6, and P1.

## 8 Applications and Area Coverage

### 8.1 Machine Learning

The GA is used in genetics-based machine learning that is an emerging area. The GAs are essential to machine learning for three factors. Firstly, it operates in discrete spaces where gradient-based techniques could not be applied. This could be used to check for rulesets, neural network structures, cellular automation machines, and many more. In this way, this could be used while stochastic high scaling and optimization algorithms could also be regarded. Secondly, these are reinforced learning techniques [45]. A specific variable, fitness, is used to assess the efficiency of the learning technique. Eventually, GA requires a population, and often whatever one needs is not a single individual, however, a community. Training in multi-agent structures is a perfect example of this [28].

### 8.2    Image Processing

The GA can be used for different digital imaging activities and the equivalent heavy pixel's algorithms. Segmentation of images is one of the key problems in the area of image processing. A lot of efficient methods are available to solve this issue. Its purpose of such a method is to divide the digital image into multiple segments based on genetic or conceptual similarities. Its purpose is to investigate the implementation of GA in the segmentation of images [29].

### 8.3    Vehicle routing problems

Multiple soft time frames, multiple depots, and heterogeneous fleet problems are also solved through the GA. Vehicle routing problems are consisting of a variety of consumers, every needing the same amount of the products to be transported. The vehicle is delivered from a single warehouse will supply the products needed and returns to the warehouse. That vehicle could bring a specific weight and could also be reduced to the distance traveled it can carry. Every consumer is permitted to visit just one vehicle. An issue is the alternative range of distribution routes that meet these criteria and have low overall costs. Throughout fact, it is also seen as analogous to reducing the total distance covered, or reducing the number of vehicles utilized, and instead of decreasing the total distance for this number of units [30].

### 8.4    Optimization Problems

The GA has been most widely used in optimized challenges where we have to maximize or reduce a given objective function value through a given set of conditions. Such a strategy will solve a problem related to optimization. The optimization raises issues of minimizing or optimizing variables with several factors that are typically subject to fairness and inequalities restrictions. This plays a pivotal role in research, marketing, and manufacturing operations. Several challenges of industrial engineering architecture are very complex and hard to overcome using traditional optimization algorithms. Within past years, GA has attracted significant attention to its capacity as an innovative optimization methodology. Depending on their usability, ease of activity, minimum specifications, and simultaneous and international viewpoint, GA has been commonly applied to a variety of issues [31].

### 8.5 Multimodal optimization

The GA becomes very strong multimodal computation methods through that we need to find solutions to several optimal problems. For technical challenges due to the physical and expense limitations, even better results achieved by a universal computation technique may not always be achieved. In these circumstances, when several approaches (local and global) are identified, the execution could be easily moved to another approach without much disruption in the design stage. It introduces a swarm multimodal optimization algorithm called cooperative human behavior [32].

### 8.6 Economics

The GA can also be used to describe various financial systems such as the cobweb system, the resolution of game theory optimization, asset pricing, etc. Several economists have started to use GAs to fix common shortcomings of conventional economic models while retaining a convenient mathematical structure. The optimization of GA happens in a way that can mimic the complexities of real decision-making issues. Also, it eliminates the need for a fixed structure to the policy problem and therefore can determine between alternate optimizations [33].

### 8.7 Neural Networks

GA also serves to practice machine learning, particularly recursive neural networks. GA is a meta-heuristic method influenced by the natural selection mechanism which belongs to the larger network of evolution optimizations. The GA is widely used to produce high-quality alternatives for computation and problem-finding by focusing on bio-inspired operations like selection, mutation, and crossover. This algorithm is a reliable approach to evolutionary optimization based on biological concepts. The population of strings describing potential solutions to problems is developed [34].

### 8.8 Parallelization

The GA seems to have very good simultaneous functionality and proves to be a very successful way to solve such problems, as well as providing a good study area. This approach is a populations-based computational optimizations technique that have been utilized to effectively train neural network systems. But, if many people making up the population, the runtime of the algorithm is always very large. Parallelism computing is a methodology that can theoretically be used to tackle this problem. Through the study explores the implementation of a parallel GA for the search for optimal specifications of artificial neural networks system. The Palatalization is accomplished using the Inter-processes communication Framework, wherein sub-populations are shared among processors during selection, mutation, and crossover [35].

### 8.9 Scheduling applications

A schedule at scheduling issues typically includes contradictory priorities based on the cost of the individual and the cost of the operator. Consumers tend to spend less time waiting, changing, and commuting by public vehicles. Technicians are involved in making money from reduced running costs for vehicles and providing a minimum number of vehicles. So far as the quality of service is involved, consumers are engaged in lower crowing whereas providers are involved with increasing income and therefore have high lead levels. Waiting time performs a significant role in scheduling planning issues. Customers are interested in integrating services with reasonable waiting times, while operators tend to provide fewer services [36]. The GA is used to solve various scheduling problems such as timetables and so on.

### 8.10 Robotics Formation

The GA can be used to design the route taken by a robotic arm from each stage to another. Various GA implementations in the area of robotic trajectory modeling have been conducted out over the last few generations. This algorithm, which is versatile public-purpose optimization algorithms, was used to produce collision-free paths for a robot with defined begin and target mutual specifications. A code system and fitness analysis are the main features of the design of the algorithm. Fitness evaluation included a measured total of the distance, time, and collisions consequences for each route [37].

### 8.11 Aircraft parametric architecture

The GA is used to design aircraft by changing variables and creating better approaches in the field of motion. A cheaper aircraft could never be the strongest, and the most effective would not be the most convenient. Under other terms, the final aircraft is still, in some context, a collaboration. Different measurements are included in the feature referred to as a quality or quantitative statistic, like value, control surfaces, mechanisms, production, fuel consumption, disturbance, and aviation dynamics, the relative weight of which depends on the expected aircraft usage [38].

### 8.12 GA in bioinformatics

GA has been used to evaluate the DNA structure using specimen spectrometric information. In recent years, the field of bioinformatics has progressed exponentially. Advancements in genomic engineering have contributed to an increase in the generation of genetic information. Such a role is challenged by the lack of awareness of genetic characteristics, as quality-altering hypotheses can quickly be implemented during algorithm construction, and the technique can easily become redundant [44]. The GAs are an evolution-inspired category of machine learning algorithms that are very promising to solve such challenges [39].

# 9      Conclusion

The GA is a probabilistic solution to optimize the problems that are modeled on a genetic evaluation process in biologically and are focused as an effective algorithm to find a global optimum solution for many types of problems. The GA is used in different artificial intelligence applications like object-oriented systems, robotics, and futuristic emerging technologies.

# 10     References

[1] Whitley D, A genetic algorithm tutorial, Statistics and computing, 1994 Jun 1;4(2):65-85.
[2] Goldberg DE, Holland JH, Genetic algorithms and machine learning, 1988.
[3] Holland JH, Genetic algorithms, Scientific American, 1992 Jul 1;267(1), 66-73.
[4] Mirjalili S, Dong JS, Sadiq AS, Faris H. Genetic algorithm: Theory, literature review, and application in image reconstruction. InNature-Inspired Optimizers 2020 (pp. 69-85). Springer, Cham.
[5] Mirjalili, Seyedali, Genetic algorithm, In Evolutionary algorithms and neural networks, pp. 43-55. Springer, Cham, 2019.
[6] Kramer, Olive, Genetic algorithm essentials, Vol. 679. Springer, 2017.
[7] S. R, R. T, A review of selection methods in genetic algorithm, Int. j. of eng. Sc. and tech., 2011 May, 3(5), 3792-7.
[8] Arabali, Amirsaman, Mahmoud Ghofrani, Mehdi Etezadi-Amoli, M. Sami Fadali, and Yahia Baghzouz. "Genetic-algorithm-based optimization approach for energy management." IEEE Transactions on Power Delivery 28, no. 1 (2012): 162-170.
[9] Mathew, T.V., 2012. Genetic algorithm. Report submitted at IIT Bombay.
[10] Tabassum M, Mathew K, A genetic algorithm analysis towards optimization solutions, International Journal of Digital Information and Wireless Communications (IJDIWC), 2014 Jan 1, 4(1), 124-42.
[11] Yang, Jinhui, Chunguo Wu, Heow Pueh Lee, and Yanchun Liang. "Solving traveling salesman problems using generalized chromosome genetic algorithm." Progress in Natural Science 18, no. 7 (2008): 887-892.
[12] Hariyadi, Putri Mutira, Phong Thanh Nguyen, Iswanto Iswanto, and Dadang Sudrajat. "Traveling Salesman Problem Solution using Genetic Algorithm." Journal of Critical Reviews, Vol 7, no. 1 (2020): 56-61.
[13] Tanweer Alam, "IoT-Fog: A Communication Framework using Blockchain in the Internet of Things", International Journal of Recent Technology and Engineering (IJRTE), Volume-7, Issue-6, 2019.
[14] Tanweer Alam, "Blockchain and its Role in the Internet of Things (IoT)", International Journal of Scientific Research in Computer Science, Engineering and Information Technology, vol. 5(1), pp. 151-157, 2019. DOI: https://doi.org/10.32628/CSEIT195137
[15] Tanweer Alam, "Internet of Things: A Secure Cloud-Based MANET Mobility Model", International Journal of Network Security, Vol. 22(3), 2020.
[16] Tanweer Alam, "Efficient and Secure Data Transmission Approach in Cloud-MANET-IoT integrated Framework", Journal of Telecommunication, Electronic and Computer Engineering (JTEC), Vol. 12 No. 1, 2020.

[17] Alam T, Benaida M. "The Role of Cloud-MANET Framework in the Internet of Things (IoT)", International Journal of Online Engineering (iJOE). Vol. 14(12), pp. 97-111. DOI: https://doi.org/10.3991/ijoe.v14i12.8338

[18] Alam, Tanweer. "Middleware Implementation in Cloud-MANET Mobility Model for Internet of Smart Devices", International Journal of Computer Science and Network Security, 17(5), 2017. Pp. 86-94

[19] Alam T, Benaida M. CICS: Cloud–Internet Communication Security Framework for the Internet of Smart Devices. International Journal of Interactive Mobile Technologies (iJIM). 2018 Nov 1;12(6):74-84. DOI: https://doi.org/10.3991/ijim.v12i6.6776

[20] Alam, Tanweer. (2018) "A reliable framework for communication in internet of smart devices using IEEE 802.15.4." ARPN Journal of Engineering and Applied Sciences 13(10), 3378-3387.

[21] Alam, Tanweer, and Mohammed Aljohani. "Design and implementation of an Ad Hoc Network among Android smart devices." In Green Computing and Internet of Things (ICGCIoT), 2015 International Conference on, pp. 1322-1327. IEEE, 2015. DOI: https://doi.org/10.1109/ICGCIoT.2015.7380671

[22] Alam, Tanweer, and Mohammed Aljohani. "An approach to secure communication in mobile ad-hoc networks of Android devices." In 2015 International Conference on Intelligent Informatics and Biomedical Sciences (ICIIBMS), pp. 371-375. IEEE, 2015. DOI: https://doi.org/10.1109/iciibms.2015.7439466

[23] Aljohani, Mohammed, and Tanweer Alam. "An algorithm for accessing traffic database using wireless technologies." In Computational Intelligence and Computing Research (ICCIC), 2015 IEEE International Conference on, pp. 1-4. IEEE, 2015. DOI: https://doi.org/10.1109/iccic.2015.7435818

[24] Alam, Tanweer, and Mohammed Aljohani. "Design a new middleware for communication in ad hoc network of android smart devices." In Proceedings of the Second International Conference on Information and Communication Technology for Competitive Strategies, p. 38. ACM, 2016. DOI: https://doi.org/10.1145/2905055.2905244

[25] Alam, Tanweer. "Fuzzy control based mobility framework for evaluating mobility models in MANET of smart devices." ARPN Journal of Engineering and Applied Sciences 12, no. 15 (2017): 4526-4538.

[26] Tanweer Alam, Mohamed Benaida. "Blockchain and Internet of Things in Higher Education." Universal Journal of Educational Research 8.5 (2020). pp 2164 - 2174. DOI: https://doi.org/ 10.13189/ujer.2020.080556

[27] Tanweer Alam, Mohamed Benaida, "Blockchain, Fog and IoT Integrated Framework: Review, Architecture and Evaluation", Technology Reports of Kansai University, Volume - 62 , Issue 02, 2020.

[28] Shapiro, Jonathan. "Genetic algorithms in machine learning." In Advanced Course on Artificial Intelligence, pp. 146-168. Springer, Berlin, Heidelberg, 1999.

[29] Jedlicka, P., Ryba, T. Genetic algorithm application in image segmentation. Pattern Recognit. Image Anal. 26, 497–501 (2016).

[30] Baker, Barrie M., and M. A. Ayechew. "A genetic algorithm for the vehicle routing problem." Computers & Operations Research 30, no. 5 (2003): 787-800.

[31] Sivanandam, S. N., and S. N. Deepa. "Genetic algorithm optimization problems." In Introduction to genetic algorithms, pp. 165-209. Springer, Berlin, Heidelberg, 2008.

[32] Cuevas, Erik, Daniel Zaldívar, and Marco Pérez-Cisneros. "A swarm optimization algorithm for multimodal functions and its application in multicircle detection." Mathematical Problems in Engineering 2013 (2013).

[33] Brooks, Arthur C. "Genetic algorithms and public economics." Journal of Public Economic Theory 2, no. 4 (2000): 493-513.
[34] Whitley, D., Starkweather, T. and Bogart, C., 1990. Genetic algorithms and neural networks: Optimizing connections and connectivity. Parallel computing, 14(3), pp.347-361.
[35] Nugroho, E.D., Wibowo, M.E. and Pulungan, R., 2017, July. Parallel implementation of genetic algorithm for searching optimal parameters of artificial neural networks. In 2017 3rd International Conference on Science and Technology-Computer (ICST) (pp. 136-141). IEEE.
[36] Shrivastava, P., Dhingra, S.L. and Gundaliya, P.J., 2002. Application of genetic algorithm for scheduling and schedule coordination problems. Journal of advanced transportation, 36(1), pp.23-41.
[37] Toogood, R., Hao, H. and Wong, C., 1995, October. Robot path planning using genetic algorithms. In 1995 IEEE International Conference on Systems, Man and Cybernetics. Intelligent Systems for the 21st Century (Vol. 1, pp. 489-494). IEEE.
[38] Marta, A.C., 2008. Parametric study of a genetic algorithm using a aircraft design optimization problem. Report Stanford University, Department of Aeronautics and Astronautics.
[39] Piserchia, Zachary. "Applications of Genetic Algorithms in Bioinformatics." PhD diss., UC Riverside, 2018.
[40] Cvjetkovic, Vladimir. "Pocket labs supported IoT teaching." International Journal of Engineering Pedagogy 8, no. 2 (2018): 32-48.
[41] Mironova, Olga, Irina Amitan, and Jüri Vilipõld. "Programming basics for beginners: Experience of the institute of informatics at Tallinn University of Technology." International Journal of Engineering Pedagogy. Vol. 7, No. 4, 2017
[42] Atoum, Issa. "A Spiral Software Engineering Model to Inspire Innovation and Creativity of University Students." International Journal of Engineering Pedagogy (iJEP) 9, no. 5 (2019): 7-23.
[43] Liao, Y.H. and Sun, C.T., 2001. An educational genetic algorithms learning tool. IEEE transactions on Education, 44(2), pp.20-pp.
[44] Tanweer Alam. mHealth Communication Framework using blockchain and IoT Technologies. International Journal of Scientific & Technology Research. Vol 9(6), 2020
[45] T. Alam "Design a blockchain-based middleware layer in the Internet of Things Architecture," JOIV : International Journal on Informatics Visualization, vol. 4, no. 1, , pp. 28 - 31, Feb. 2020. https://doi.org/10.30630/joiv.4.1.334
[46] Rajsingh, Elijah Blessing, Jey Veerasamy, Amir H. Alavi, and J. Dinesh Peter, eds. Advances in Big Data and Cloud Computing. Vol. 645. Springer, 2018.